\documentclass[nofootinbib,prd,twocolumn,10pt]{revtex4-1}
\pdfoutput=1
\usepackage{amsmath,amssymb}
\usepackage{epsfig}
\usepackage{graphicx}
\usepackage[usenames,dvipsnames]{color}
\usepackage{subfigure}
\usepackage{slashed}
\usepackage{float}
\usepackage[colorlinks,citecolor=blue]{hyperref}
\usepackage{color}
\usepackage{multirow}
\usepackage{xcolor}
\usepackage{soul}
\usepackage[normalem]{ulem}
\begin{document}
\begin{flushright}
PI-UAN-2022-715FT
\end{flushright} 

\title{New $W$-Boson mass in the light of doubly warped braneworld model}
\author{Basabendu Barman$^{1}$,\,Ashmita Das $^{2}$,\,Soumitra Sengupta$^{2}$}

\email{basabendu88barman@gmail.com}
\email{ashmita.phy@gmail.com}
\email{tpssg@iacs.res.in}

\affiliation{\,Centro de Investigaciones, Universidad Antonio Nari\~{n}o\\Carrera 3 este \# 47A-15, Bogot{\'a}, Colombia$^{1}$}

\affiliation{School of Physical Sciences, Indian Association for the Cultivation of Science, Kolkata 700032, India$^{2}$}

\begin{abstract}
The recent observation by CDF collaboration has disclosed a modification in the mass of the $W$ boson. In this regard we show that this modification in the mass of the $W$ boson can be well explained in the background of a 6-dimensional warped geometry model, where the double warping is associated with the two extra spatial dimensions. We consider that all the Standard Model fields are residing in the bulk, where the bulk Higgs field gives rise to the spontaneous symmetry breaking in the 6-dimensional spacetime.  
Allowing a little hierarchy between the two moduli we exactly obtain the observed mass for the $W$ boson, which is identified as the lowest lying Kaluza-Klein mass mode of the bulk $W$ boson on the $(3+1)$ dimensional visible brane. The essential feature of the 5-dimensional Randall-Sundrum scenario such as the resolution of the gauge hierarchy problem without introducing any intermediate scale between the Planck and the TeV scale, remains intact.
\end{abstract}
\maketitle

Recently, the $W$ boson mass has been accurately measured by the CDF II detector with 8.8 $\text{fb}^{-1}$ data accumulated at the Fermilab Tevatron collider with a center of mass energy of $\sqrt{s}=1.96$ TeV~\cite{CDF:2022hxs}. The newly measured $W$ boson mass is

\begin{equation}
M_W^\text{CDF} = 80,433.5 \pm 9.4\,\text{MeV}.    
\end{equation}

From the Standard Model (SM), on the othr hand, the expected $W$ mass is

\begin{equation}
M_W^\text{SM} = 80,357\pm 6\,\text{MeV}\,,    
\end{equation}

\noindent making the new CDF measurement larger than the SM expectation by the amount $\Delta M_W=76$ MeV, which is about 7$\sigma$ standard deviation away from the SM prediction. Before the CDF run-II result, the world average of $M_W$ measurements was just 1.8$\sigma$ standard deviation from the SM expectation. In order to explain this humongous discrepancy, numerous new physics (NP) scenarios have been discussed in, for example, Refs.~\cite{Fan:2022dck,Liu:2022jdq,Athron:2022isz,Song:2022xts,Cheung:2022zsb,Endo:2022kiw,Han:2022juu,Ahn:2022xeq,Perez:2022uil,Kawamura:2022uft,Kanemura:2022ahw,Nagao:2022oin,Mondal:2022xdy,Zhang:2022nnh,Carpenter:2022oyg,Popov:2022ldh,Arcadi:2022dmt,Chowdhury:2022moc,Borah:2022obi,Du:2022fqv,Ghorbani:2022vtv,Asadi:2022xiy,Bhaskar:2022vgk,Babu:2022pdn,Ghoshal:2022vzo,Arias-Aragon:2022ats,Sakurai:2022hwh,Gu:2022htv,Batra:2022org,Baek:2022agi,Borah:2022zim,Heeck:2022fvl,Addazi:2022fbj,Cheng:2022aau,Crivellin:2022fdf,Lee:2022gyf,Batra:2022pej,Benbrik:2022dja,Cai:2022cti,Zhou:2022cql,Zhu:2022tpr,Wang:2022dte}. All of these models incorporate new fields beyond the SM that contributes to the so called $\rho$-parameter making it compatible with the CDF measurement, which is close to unity otherwise. Certainly this is an obvious way of explaining the deviation from SM prediction. Although there can be numerous viable interesting particle physics models that can do the job, but this can also be motivated from a different perspective. Such instances, however, are only a handful. For example, in~\cite{Heckman:2022the} the author has discussed how a stringy particle physics scenario can generate a shift in the $\rho$ parameter in accordance with the CDF measurement.

In this manuscript we aim to shed light on this observation, in the background of a 6-dimensional doubly warped geometry model which was introduced in~\cite{PhysRevD.76.064030}. This 6-dimensional doubly warped geometry with flat 3-branes is an extension of the original Randall-Sundrum (RS) scenario~\cite{PhysRevLett.83.3370} to more than one extra dimensions \footnote{For reviews on the RS model, see, for example~\cite{Kim:2003pc,Csaki:2004ay,Csaki:2015xpj}.}. In the RS model, the SM fields are assumed to be located on the TeV brane while only gravity propagates in the bulk \footnote{Phenomenology of bulk SM fields in the warped geometry model has been explored in \cite{Goldberger:1999wh,Davoudiasl:2000wi,Pomarol:1999ad,Grossman:1999ra,Chang:1999nh}.}.  
It was shown that the 5-dimensional RS model with bulk gauge and Higgs fields, and the spontaneous symmetry breaking taking place in the bulk, fails to generate the SM $W$ and $Z$ boson masses $\lesssim 100\,\rm GeV$ on the visible/TeV/SM brane. Moreover, the first excited state in the Kaluza-Klein (KK) mass tower of an abelian bulk gauge boson exhibits a large coupling with fermions, which  constrains the mass of this state by such an amount that the model may survive the direct search bound at the Fermilab Tevatron and the precision electroweak constraints. However, the mass of the first excited KK mode of this abelian bulk gauge boson appears to be substantially lowered than the bounds proposed above. One may try to resolve these issues by adjusting the bulk parameters of RS model, however that would jeopardize the resolution of the gauge hierarchy problem, which was the prime motive of such a warped geometry model. In Refs.~\cite{Davoudiasl:2000wi,Chang:1999nh,Huber:2000fh, Davoudiasl:2005uu, Archer:2010hh} both of these problems have been discussed in detail.

These problems of bulk 5-dimensional RS scenario has been successfully addressed in the background of a 6-dimensional doubly warped model with flat $(3+1)$ dimensional branes \cite{Das:2011fb}. As shown in \cite{Das:2011fb}, this 6-dimensional RS setup is capable of accommodating massive gauge bosons of mass $\sim 100$ GeV on the $(3+1)$ dimensional visible brane. This motivates us to explain the newly measured $W$ boson mass in the present framework. In this setup, we show, it is indeed possible to obtain the lowest lying KK mode of the bulk $W$ boson on the $(3+1)$ dimensional visible brane complying with the CDF measurement, without harming the main essence of the RS scenario. Our goal is to establish that this new observation can be a potential hint towards the higher dimensional theories with more than one extra dimensions, as already mentioned that theories with one extra dimension are very tightly constrained from the existing collider limits~\cite{Kubota:2014mma,ATLAS:2020fry}. 

The paper is organized as follows. In Sec.~\ref{sec:model} we briefly discuss the 6-dimensional RS model illustrating how  masses of the SM gauge bosons can be generated in this set-up, we then discuss our results in Sec.~\ref{sec:result} and finally conclude in Sec.~\ref{sec:concl}.  

\section{The Formalism}
\label{sec:model}
\begin{figure}[H]
    \centering
    \includegraphics[scale=0.45]{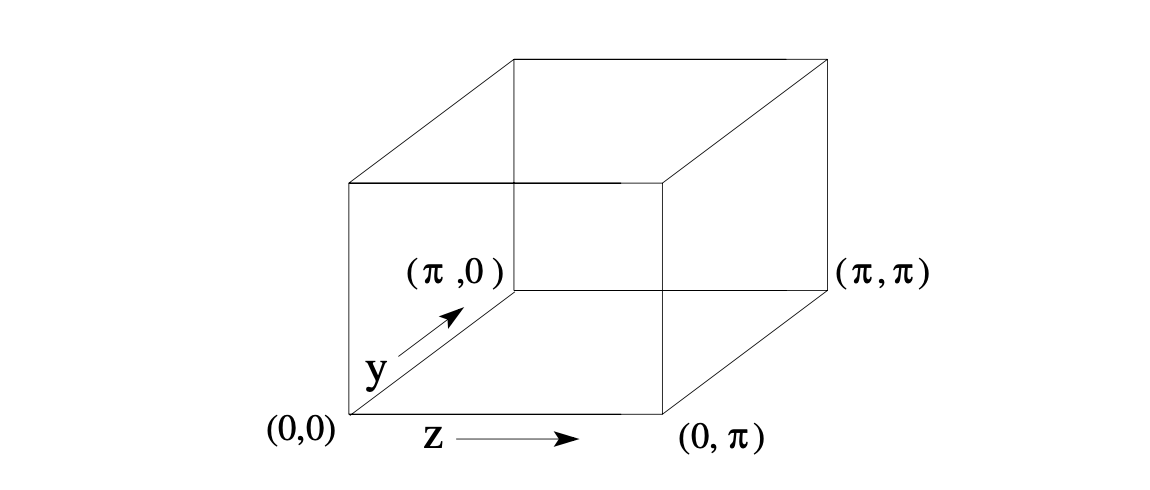}
    \caption{The 6-dimensional doubly warped geometry}
    \label{fig:6d}
\end{figure}

This 6-dimensional doubly warped model contains two extra spatial dimensions along with the usual 4 noncompact dimensions. The double warping is associated with the two extra spatial dimensions which are denoted by the angular coordinates ($y,\,z$). As similar to the 5-dimensional RS scenario, here the extra dimensions are considered to be $Z_2$ orbifolded and the entire 6-dimensional spacteime can be portrayed as : $[M^{(1,3)}\times S^1/Z_2] \times S^1/Z_2$. The spacetime metric corresponding to this geometry can be conceived as following,  
%
\begin{equation}
ds^2 = b^2(z)\left[a^2(y)\eta_{\mu\nu}dx^\mu dx^\nu +R_y^2dy^2\right]+r_z^2dz^2\,.
\label{E:metric}
\end{equation}
\noindent Here, $R_y,r_z$ depict extra dimensional moduli corresponding to the extra dimensions $y,z$ respectively. $x^\mu$, $\mu=0,\cdots,3$ symbolises the usual 4 noncompact dimensions. The signature of the above metric has taken to be $\{-,+,+,+\}$. $a(y)$ and $b(z)$ represent the warp factors along the extra dimensions $y$ and $z$
respectively. In this setup one can write the 6-dimensional bulk, $(4+1)$ and $(3+1)$ dimensional brane actions as following  \cite{PhysRevD.76.064030}, 
\begin{align}
& S = S_6+S_5+S_4
\nonumber \\&
S_6 =\int d^4xdydz\sqrt{-g_6}\,(R_6-\Lambda), \quad
\nonumber \\&
S_5=\int d^4xdydz\,\sqrt{-g_{5}}\,[V_1\delta(y)+V_2\delta(y-\pi)]
\nonumber\\&
     +\int d^4xdydz \,\sqrt{-g_{5}}\,[V_3\delta(z)+V_4\delta(z-\pi)]\nonumber\\
    & S_4=\int d^4xdydz\,\sqrt{-g_{\rm vis}}\,[\mathcal{L}-\hat{V}].
\end{align}
\noindent Here, $V_{1,2}$ and $V_{3,4}$ symbolise the brane tensions of the $(4+1)$-dimensional branes located at $y=0,\pi$ and $z=0,\pi$, respectively. The 6-dimensional cosmological constant is denoted by $\Lambda$.
With the above metric ansatz the 6-dimensional warped spacetime acquires a box like shape as shown in the Fig.\ref{fig:6d}, where the walls of the box depict $(4+1)$ dimensional branes and the intersections of $4$ branes form $(3+1)$ dimensional branes.
One can indeed solve the 6-dimensional Einstein's equations, while using the metric ansatz as in Eq.(\ref{E:metric}) and obtain the functional form of the warp factors as below \cite{PhysRevD.76.064030},
\begin{eqnarray}
a(y) &=& \exp(-c|y|), \quad b(z) = \frac{\cosh(kz)}{\cosh(k\pi)}
\nonumber \\
c&\equiv & \frac{R_yk}{r_z\cosh(k\pi)},\quad k\equiv r_z\sqrt{
\frac{-\Lambda}{10M_P^4}}.
\label{E:sol}
\end{eqnarray}
\noindent 
Here, $M_P$ denotes the Planck mass scale. It can be seen  that the metric on the 3-brane located at ($y=0, z=\pi$) suffers no warping and thus can be identified with the Planck brane. The $(3+1)$-dimensional visible brane can be located anywhere among the following options, $(y,z)=(0,0),(\pi,0)$ or $(\pi,\pi)$ depending on the values of $(k,\,c)$. It is well explained in \cite{PhysRevD.76.064030} that considering our Universe/visible brane possesses the lowest natural energy scale than the other $(3+1)$-dimensional branes, the location of the visible brane can be set uniquely at $(y=\pi, z=0)$. The warp factors at this location {\it i.e,}
\begin{equation}
a\,(y=\pi)\,b\,(z=0)=f =\frac{\exp(-c\pi)}{\cosh(k\pi)}\,, 
\label{E:supp}
\end{equation}
yields largest suppression on the brane. In order to resolve the gauge hierarchy problem  similar to the 5-D RS scenario, the required suppression on the visible brane turns out to be $f \sim  \mathcal{O}(10^{-16})$. At this stage couple of comments are in order : (i) $f$ can be obtained as $\sim 10^{-16}$ for different choices of $(c,k)$, (ii) being motivated form the original RS scenario, we do not allow any intermediate scales between the Planck and the TeV scale in the present work. Therefore we take both the moduli $R_y$ and $r_z$ to be $\mathcal{O}$ (Planck length), which implies two possibilities for the values of $(c,k)$ such as large $c$, small $k$ or large $k$, small $c$, as can be seen from the Eq.~\eqref{E:sol}.   
Recently it was shown in \cite{Bhaumik:2022xtd} that these two moduli can be stabilized to the desired values by a 6-dimensional bulk scalar field, which possesses nonzero VEVs on the 3-branes. This moduli stabilization has also been addressed in the context of string theory while using the fluxes \cite{Balasubramanian:2005zx}. 
This 6-dimensional model can successfully address the gauge hierarchy problem without bringing in any intermediate and hierarchical scale in the theory.  
 This model has many phenomenological promises, for example  the observed mass hierarchy among the SM fermions can be explained in the background of this 6-dimensional model \cite{PhysRevD.76.064030}. Moreover in this model the mass of the first excited graviton KK mode on the visible brane becomes significantly large while the coupling of the same with the SM fields turns out to be heavily suppressed in comparison to the original RS scenario. This may explain the invisibility of the graviton KK modes in the LHC \cite{Arun:2014dga}. 
\subsection*{The SM gauge boson mass in the bulk}
It was shown in \cite{Das:2011fb} that the present model can accommodate Higgs mechanism in the bulk of the 6-dimensions and the vev of the Higgs field will be of the order of Planck scale. The vev of the Higgs field contributes to generate the bulk mass for the gauge field. The invariant action can be written as

\begin{align}
&S_G = \int d^4xdydz\sqrt{-G}\Bigl(-\frac{1}{4}G^{MK}G^{NL}F_{KL}F_{MN}
\nonumber\\&
-\frac{1}{2}M^2G^{MK}A_MA_K\Bigr)\,,
\label{E:SG}
\end{align}

\noindent where $M$ is the bulk mass $\sim M_P$ and $G={\rm det}(G_{AB})$ is the determinant of the metric $G_{AB}$ which is given in Eq. (\ref{E:metric}).
$F_{MN}=D_M A_N-D_N A_M$ is the gauge field strength.
Exploiting the gauge symmetry we can choose the gauge condition where $A_4=A_5=0$. 
While the above action is written
for a general bulk gauge field ($A_{M}$), for the interest of our present work we write the above action in terms of all the gauge bosons
as following, 
\begin{eqnarray}
&S_G = \int d^4x\,dy\,dz\sqrt{-G}\bigg[\bigg\{-\frac{1}{4} W_{MN} W^{MN}+M_{W}^{2}\,W^{+}\, W^{-}\bigg\}\nonumber\\
&- \frac{1}{4} B_{MN} B^{MN}+\frac{1}{2} M_{Z}^{2} Z_{M}^{2}+\frac{1}{2}M_{A}^{2} A_{M}^{2}\bigg]
\label{E:SG:1}
\end{eqnarray}
 Here $W^{\pm}$, $W_{MN}$ depict the 6-dimensional massive $W$ boson and its field strength tensor respectively.  $B_{M},\,B_{MN}$ represents the fourth gauge field and its field strength in the 6 dimensions. $M_{W}, \,M_{Z}$ and $M_A$ symbolise the acquired mass for bulk $W, \,Z$ bosons and $U(1)$ field  via spontaneous symmetry breaking in the bulk.
It is worth to be mentioned at this stage that one needs to put $M_{A}=0$ in order to obtain the lowest KK mass mode for the bulk $U(1)$ field as zero on the $(3+1)$ dimensional visible brane. This lowest lying KK mode of the bulk $U(1)$ field is interpreted as SM photon on the same \cite{Das:2011fb}. 
It can be seen that Eq.(\ref{E:SG:1}) depicts the explicit form of the 6-dimensional action corresponding to the bulk $W$ boson which plays the central role in the present work .

For further analysis we choose to work with the generic 6-dimensional gauge field $A_M$ and the action in Eq.(\ref{E:SG}).
We consider the KK mode decomposition of such gauge fields as, 
\begin{equation}
A_\mu = \sum_{n,p}A^{(n,p)}_\mu(x)\xi_n(y)\chi_p(z)/\sqrt{R_yr_z}\,,
\label{E:KKgau}
\end{equation}
where $A^{(n,p)}_\mu(x)$ are the KK modes on the visible brane which hold two indices ($n,p$) corresponding to the two extra dimensions. $\xi_n(y)$ and $\chi_p(z)$ represent the KK wave functions along the two extra dimensions $(y,z)$ respectively. In order to obtain the $(3+1)$-dimensional effective action we use Eq.(\ref{E:KKgau}) in the action (\ref{E:SG}) and integrate over the two extra dimensions $y,z$. It is natural to claim that the outcome of the above integration will eventually lead us to the effective action of the form 
\begin{equation}
\sum_{n,p}-\frac{1}{4}F_{\mu\nu}^{(n,p)}F^{(n,p)\mu\nu}-\frac{1}{2}
m_{n,p}^2A_{\mu}^{(n,p)}A^{(n,p)\mu}\,.
\label{kk_effective}
\end{equation}
In the above $m_{n,p}$ denotes the mass of the KK modes on the $(3+1)$-dimensional brane. To obtain the Eq. (\ref{kk_effective}) the KK wave functions should obey the orthonormality conditions and the eigenvalue equations as , 
\begin{equation}
\int dy~\xi_n(y)\xi_{n^\prime}(y) = \delta_{nn^\prime},
\quad
\int dz~b(z)\chi_p(z)\chi_{p^\prime}(z) = \delta_{pp^\prime}\,.
\label{E:gnorm}
\end{equation}
\begin{eqnarray}
\frac{1}{R_y^2}\partial_y(a^2\partial_y\xi_n)-m_p^2a^2\xi_n &=& -m_{n,p}^2\xi_n,
\nonumber \\
\frac{1}{r_z^2}\partial_z(b^3\partial_z\chi_p)-M^2b^3\chi_p &=& -m_{p}^2\chi_p\,,
\label{E:KKyz}
\end{eqnarray}
\noindent 
where $m_p$ is designated to be a mass parameter and determined by solving the equation for $\chi_p(z)$. We take an approximation {\it i.e} $b(z)\sim\exp[-k(\pi-z)]=\exp[-k\tilde{z}]$ and write  $\tilde{\chi}_p(z)=\exp(-3k\tilde{z}/2)\chi_p(z)$ in the last part of the Eq. (\ref{E:KKyz}) in order to solve the equation for $\chi_p (z)$. This yields, 

\begin{equation}
z_p^2\frac{d^2\tilde{\chi}_p}{dz_p^2}+z_p\frac{d\tilde{\chi}_p}{dz_p}
+(z_p^2-\nu_p^2)\tilde{\chi}_p=0\,,
\end{equation}
\noindent where $z_p=\frac{m_p}{k^\prime}\exp(k\tilde{z})$ and $\nu_p^2=\frac{9}{4} +\left(\frac{M}{k^\prime}\right)^2$. Here, $k^\prime=k/r_z$. 
Solving the above equation one gets the Eq.(\ref{E:Zeigen}),  
\begin{equation}
\chi_p(z)=\frac{1}{N_p}\exp\left(\frac{3}{2}k\tilde{z}\right)\,\left[J_{\nu_p}(z_p)+b_pY_{\nu_p}(z_p)\right]\,,
\label{E:Zeigen}
\end{equation}
\noindent where $\nu_p$ is the order of the bessel function and $N_P$, $b_p$ are some constants.
Imposing the boundary conditions that the $\chi_p(z)$ is continuous at the orbifold fixed points $z=0,\pi$, we obtain
\begin{equation}
3J_{\nu_p}(x_{\nu_p})+x_{\nu_p}(J_{\nu_p - 1}(x_{\nu_p}) - J_{\nu_p + 1}(x_{\nu_p})) = 0\,,
\label{E:Zeq}
\end{equation}
\noindent which decides the mass spectrum $m_p$. Here,  $x_{\nu_p}=\frac{m_p}{k^\prime}\exp(k\pi)$. For a fix value of $k$, using Eq. (\ref{E:Zeq}) and  $x_{\nu_p}=\frac{m_p}{k^\prime}\exp(k\pi)$, one obtains the value for $\frac{m_p}{k^\prime}$.
At this stage we further reparametrize 
$\tilde{\xi}_n
=\exp(-c|y|)\xi_n$ in the first equation of (\ref{E:KKyz}), which yields, 
\begin{equation}
y_n^2\frac{d^2\tilde{\xi}_n}{dy_n^2}+y_n\frac{d\tilde{\xi}_n}{dy_n}
+(y_n^2-\nu_n^2)\tilde{\xi}_n=0\,,
\end{equation}
\noindent where $y_n=\frac{m_{n,p}}{k^\prime}\exp(c|y|)\cosh(k\pi)$ and
$\nu_n^2=1+\left(\frac{m_p}{k^\prime}\right)^2\cosh^2(k\pi)$. As similar to the $\chi_p(z)$, we obtain the solution for $\xi_n(y)$ as below, 
\begin{equation}
\xi_n(y)=\frac{1}{N_n}\exp(c|y|)
\left[J_{\nu_n}(y_n)+b_nY_{\nu_n}(z_n)\right]\,,
\label{E:Yeigen}
\end{equation}
\noindent where $\nu_n$ is the order of the Bessel function and $N_n$, $b_n$ are some constants.
Implementing the similar boundary condition on $\xi_n(y)$ as $\chi_p (z)$, one gets, 
\begin{equation}
J_{\nu_n}(x_{\nu_n})+x_{\nu_n}(J_{\nu_n - 1}(x_{\nu_n}) - J_{\nu_n + 1}(x_{\nu_n}))/2 = 0,
\label{E:Yeq}
\end{equation}
where,
\begin{equation}
x_{\nu_n}=\frac{m_{n,p}}{k^\prime}\exp(c\pi)\cosh(k\pi)\,.
\label{E:xn}
\end{equation}
and $m_{n,p}$ signifies the induced KK mass for the bulk gauge fields on the $(3+1)$-dimensional visible brane. It is worth mentioning at this stage that in this 6-dimensional model, we have a freedom of two moduli scales as opposed to one modulus in the 5-D RS scenario. Allowing a little hierarchy between these moduli, it was shown in~\cite{Das:2011fb} that the right magnitude for $W$ and $Z$ boson masses can be achieved from the KK modes of the massive bulk gauge fields where the spontaneous symmetry breaking is triggered by bulk Higgs. 
In the context of the present work where $m_{n,p}$ represent the KK mass modes of the bulk $W$ boson on $(3+1)$ dimensional visible brane, we consider the mass mode corresponding to $n=p=1$ {\it i.e} $m_{1,1}$ which portrays the mass of the SM $W$ boson. This represent  the lowest lying KK mode of the bulk $W$ boson. 
\section{Results and Discussions}
\label{sec:result}

\begin{figure}
    \centering
    \includegraphics[scale=0.4]{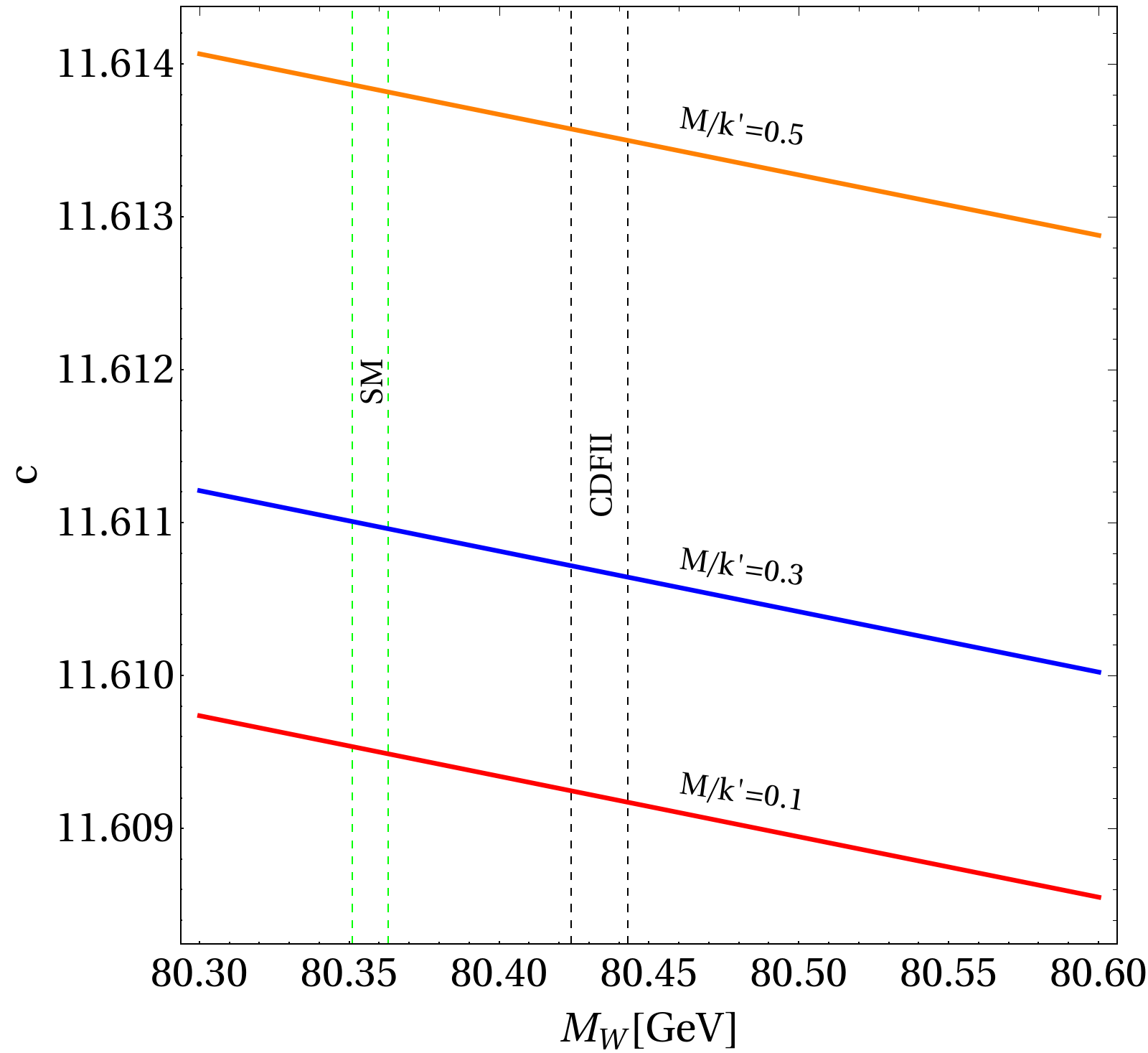}
    \caption{Variation of $c$ with the mass of the lightest KK mode of bulk $W$ boson on the $(3+1)$ dimensional visible brane is shown. We fix $k=0.25$. The vertical black dashed lines correspond to $M_W^\text{CDF} = 80,433.5 \pm 9.4\,\text{MeV}$, while the vertical green dashed lines are the SM predictions.}
    \label{fig:mw-rs}
\end{figure}

In order to examine the phenomenological merits of this model in the light of the recent observation at CDF II, we make couple of comments as follows,
\begin{itemize}
    \item 
    The 6-dimensional warped geometry model is a possible extension to the 5-D RS scenario. Thus in our entire analysis we keep intact the main success of the original RS model such as the resolution of the gauge hierarchy problem. This compels us to maintain the suppression factor on the SM brane $\mathcal{O}(10^{-16})$.
    \item
    Referring to the discussion below the Eq.(\ref{E:supp}), we ensure not to have a large hierarchy in the moduli $R_y$ and $r_z$. Thus we take $c$ and $k$ differ by just one order, which leads to an order one hierarchy between the two moduli. Therefore both the moduli $R_y$ and $r_z$ remain close to the Planck length. 
\end{itemize}
Now from Eq.(\ref{E:xn}) we aim to examine the variations in the extra dimensional parameters such as $c$ or $k$ with respect to the mass of the SM $W$ boson where our prime focus will be on the mass range as proposed by CDF II. Thus for a bulk $W$ boson, at first we take certain values for $M/k'$ without allowing much hierarchy in the bulk parameters and subsequently fix the parameters $k=0.25$ and $k'=2\times 10^{17}$ GeV. This enables one to write the parameter $c$ as a function of $m_{1,1}$ which in turn generates the plot of $c$ vs $m_{1,1}$ as depicted in Fig.\ref{fig:mw-rs}.

Fig.~\ref{fig:mw-rs} depicts the variation of the parameter $c$ which controls the warping along the $y$ direction with the mass of the lowest lying KK mode of bulk $W$ boson on $(3+1)$ dimensional visible brane. The mass range of $W$ boson is varied due to the recently observed CDF II data where the CDF II-observed mass window is shown via the shaded vertical region.
This plot suggests that it is indeed possible to accommodate the newly observed $W$ mass within the bulk 6-D RS scenario without jeopardizing the resolution to the hierarchy problem. 
Here we emphasize that the choice of  $k=0.25$ and the set of  values of $c$ in the above plot yields $R_y/r_z\sim\mathcal{O}(10)$. This implies that the observed mass of the $W$ boson can be achieved on the $(3+1)$ dimensional visible brane, without introducing any intermediate scale in between the Planck and the TeV scale. This choice of the parameters is motivated by~\cite{Bhaumik:2022xtd}, where by generalizing the Goldberger-Wise stabilization mechanism in 6 dimension, it has been shown that within the range of $k \ni 0.1 - 0.6$, the stability of both the moduli of the present model can be achieved by a 6-dimensional bulk scalar field without requiring  any unnatural fine tuning. Thus the relevant moduli parameters in the present work are chosen following the requirement of moduli stabilization.
it was further shown in \cite{Das:2011fb} that this 6-dimensional model not only generates the correct masses for the SM gauge bosons but also reproduces  the appropriate  gauge couplings and the KK gauge boson masses which  satisfy the precision electroweak constraints and respect the Tevatron bounds for the same choice of parameters as considered here.

\section{Conclusions}
\label{sec:concl}

The new measurement of the $W$ mass reported by the
CDF collaboration features a large deviation from the
theoretical expectation in the SM, but is also hugely different from previous measurements made at experiments like LEP, the Tevatron and LHC. This has created quite a ripple by hinting towards numerous scopes for beyond the Standard Model (SM) physics. Albeit more careful analyses are needed in order to confirm this result, nonetheless one can always ask what kind of physics beyond the SM can be accommodated within this window. As numerous particle physics models have been already been proposed to explain this anomaly, however there can be hints from more exotic physics like extra dimension or string theory. 
With this motivation, in the present work we prescribe a mechanism within the warped extra dimensional scenario which can accommodate a massive $W$  boson, complying with the CDF II result. We illustrate that the newly observed $W$ boson mass can be explained in the background of a 6-dimensional warped geometry model, where all the SM fields being located in the bulk. This model possesses a larger parameter space due to the presence of an extra modulus than that of the 5-dimensional RS model. We show that by setting one of the moduli approximately two orders smaller than the Planck scale, one acquires the mass of the lowest lying mode of the bulk $W$ boson on the visible brane as exactly observed in CDFII. The choice of the parameters do not contradict the main spirit of the original RS model. Throughout the analysis the warp factor on visible brane is maintained to be $\sim 10^{-16}$ that ensures the resolution of the gauge hierarchy problem. Moreover, in the background of this model the gauge couplings and the masses of the KK gauge fields satisfy the precision electroweak constraints and also respect the Tevatron bounds~\cite{Das:2011fb}, which reveals the additional successes of this 6 dimensional warped geometry model over the original RS scenario. 
The present model can be extended to even higher dimensional spacetime with more than two warped extra dimensions~\cite{PhysRevD.76.064030}. Consequently, our findings will also be generalized to the larger dimensions. Looking into the Eq.~\eqref{E:xn}, it can be realised that, with the inclusion of more than two extra dimensions, the effective parameters on the visible brane, such as the masses of the KK modes of the bulk gauge fields, turn up to be the function of more parameters (like $(c, k)$) associated with the extra dimensions. Therefore adjusting the moduli appropriately, the associated parameters of the different gauge bosons can be adjusted to the desired values, while satisfying the precision electroweak constraints and the Tevatron bounds.  
Thus the background warped geometry is responsible to produce the observed phenomena on the visible/TeV brane while at the same time addressing the well-known naturalness problem. This suggests  that the phenomenological implications  of this model are worth studying in the light of the recent data from the CDFII.
\section*{Acknowledgements}
BB received funding from the Patrimonio Autónomo - Fondo Nacional de Financiamiento para la Ciencia, la Tecnología y la Innovación Francisco José de Caldas (MinCiencias - Colombia) grant 80740-465-2020. This project has received funding /support from the European Union's Horizon 2020 research and innovation programme under the Marie Sklodowska-Curie grant agreement No 860881-HIDDeN. AD would like to thank Indian Association for the Cultivation of Science, Kolkata, for providing an academic visit, during which the present work is accomplished.   

\appendix

\bibliography{Bibliography}

\end{document}